\documentclass[nofootinbib, twocolumn,secnumarabic,amssymb, nobibnotes, aps, pra,superscriptaddress,longbibliography,floatfix,10pt]{revtex4-2}

\usepackage{amsmath}
\usepackage{longtable}
\usepackage{makecell}
\usepackage{graphicx}
\usepackage{float}
\usepackage{booktabs}
\usepackage{url}
\usepackage{mathtools}

\usepackage{color}
\usepackage{multirow}
\usepackage{diagbox}
\usepackage[colorlinks,linkcolor=blue,anchorcolor=blue,citecolor=blue,urlcolor=blue]{hyperref}
\usepackage{microtype}
\newsavebox{\mybox}

\graphicspath{{Figures/}}

\usepackage{array}
\makeatletter
\newcommand{\thickhline}{%
    \noalign {\ifnum 0=`}\fi \hrule height 1pt
    \futurelet \reserved@a \@xhline
}
\newcolumntype{"}{@{\hskip\tabcolsep\vrule width 1pt\hskip\tabcolsep}}
\makeatother
\usepackage{lipsum}

\begin{document}
\raggedbottom

\title{Photon-conserving Raman soliton attractors in focusing and defocusing Kerr media}

\author{Weiye Huang}
\affiliation{State Key Laboratory of Information Photonics and Optical Communications, Beijing University of
    Posts and Telecommunications, Beijing 100876, China}
    
\author{Junhong Yang}
\affiliation{KEY $\&$ Core Technology Innovation Institute of the Greater Bay Area, Guangzhou 510725, China}

\author{Tao Sun}
\affiliation{KEY $\&$ Core Technology Innovation Institute of the Greater Bay Area, Guangzhou 510725, China}

\author{Qian Gao}
\affiliation{KEY $\&$ Core Technology Innovation Institute of the Greater Bay Area, Guangzhou 510725, China}
    
\author{Peilong Yang}
\affiliation{Laboratory of Infrared Materials and Devices,
    The Research Institute of Advanced Technologies, Ningbo University, Ningbo 315211, China}

\author{Jintao Fan}
\affiliation{Ultrafast Laser Laboratory, Key Laboratory of Optoelectronic Information Science and Technology of the Ministry of Education, School of Precision Instruments and Opto-electronics Engineering, Tianjin University, Tianjin 300072, China
}

\author{G{\"u}nter Steinmeyer}
\affiliation{Institut f{\"u}r Physik, Humboldt-Universit{\"a}t zu Berlin, Newtonstra{\ss}e 15, 12489 Berlin, Germany}
\affiliation{Max Born Institute for Nonlinear Optics and Short Pulse Spectroscopy, Max-Born-Stra{\ss}e 2a, 12489 Berlin, Germany}

\author{Jinhui Yuan}
\email{yuanjinhui81@bupt.edu.cn}
\affiliation{State Key Laboratory of Information Photonics and Optical Communications, Beijing University of Posts and Telecommunications, Beijing 100876, China}

\author{Chao Mei}
\email{meichao@nbu.edu.cn}
\affiliation{Department of Physics, School of Physical Science and Technology, Ningbo University, Ningbo 315211, China}
\date{\today}

\begin{abstract}
\noindent The sign of the Kerr nonlinear coefficient has long been regarded as irrelevant to the direction of the Raman-induced soliton self-frequency shift. Yet the standard generalized nonlinear Schrödinger equation (GNLSE) predicts a frequency shift that depends on the sign of the nonlinearity, which leads to an unphysical blue shift in the defocusing case. We resolve this inconsistency by deriving the time-domain photon-conserving GNLSE (pcGNLSE) from its established frequency-domain counterpart. The derivation identifies additional carrier-frequency terms in the Kerr and Raman operators. The resulting energy balance predicts a decrease in localized optical pulse energy, independent of the sign of the Kerr coefficient or its frequency slope.
Applying the method of moments, we derive five-parameter evolution equations for bright and dark solitons and obtain constant-peak-power and constant-blackness conditions that identify Raman-soliton attractors in both focusing and defocusing Kerr media. These analytical conditions relate pulse parameters directly to dispersion, nonlinear dispersion, and Raman corrections, providing a physically consistent framework for designing Raman-soliton dynamics in emerging semiconductor waveguide and microresonator platforms.
\end{abstract}
\maketitle

\section{Introduction} \label{sec:intro}

\noindent Optical solitons exist only when group-velocity dispersion (GVD) is precisely canceled by the nonlinear phase shift from the Kerr effect~\cite{Hasegawa1973,Mollenauer1980}. In real optical fibers and waveguides, however, higher-order effects perturb this ideal balance, and stimulated Raman scattering is particularly consequential for femtosecond and picosecond pulses. Intrapulse Raman gain preferentially amplifies the red spectral wing at the expense of the blue wing, producing a continuous downshift of the soliton center frequency~\cite{Gordon1986,Mitschke1986} known as the soliton self-frequency shift. This effect has been exploited in wavelength-tunable soliton sources~\cite{Wang2025,Zhang2025}, beam self-cleaning~\cite{Chen2026}, coherent Raman microscopy~\cite{Xie2025} and all-optical signal processing~\cite{Mei2019}.

The standard theoretical framework for all these phenomena is the generalized nonlinear Schr\"{o}dinger equation (GNLSE)~\cite{Blow1989,Agrawal2019}, which typically includes higher-order dispersion (TOD), self-steepening (SS), and a delayed Raman response. Within this framework, the soliton attractor concept has emerged as a useful organizing principle. Rather than seeking a strictly stationary solution, one asks whether a broad class of initial conditions converges to a manifold of pulse shapes~\cite {Akhmediev1997,Soto-Crespo2005,Baals2021}. For the GNLSE, bright-soliton attractors have been identified in the anomalous-dispersion regime, where the interplay between the Raman redshift, SS-induced broadening, and dispersion stabilizes the peak power against perturbations~\cite{Rahaman2025,Biancalana2004}. Formally, bright soliton attractors can also occur in the normal-dispersion regime when the effective Kerr nonlinearity is defocusing. Negative Kerr nonlinearities can occur in material systems that also exhibit strong two-photon absorption~\cite{SheikBahae:91}, including two-dimensional materials~\cite{Margulis2018} and nanoparticle-doped media~\cite{Bose2016}. In such systems, however, dissipative effects can obscure Raman-driven soliton dynamics. A much more tangible situation arises in cascaded $\chi^{(2)}$ interaction \cite{Stegeman1996Cascading}. This effect has been exploited to directly support soliton mode-locking in the normal-dispersion regime \cite{Phillips:15} and to achieve high conversion efficiencies in THz generation schemes \cite{Jolly19}. Using cascaded difference-frequency generation, strong Raman-induced red-shifts have been experimentally observed \cite{Ravi20}. 

In contrast, extending the conventional GNLSE to negative nonlinearity reverses its Raman term and predicts a spectral blueshift. This behavior originates from two assumptions in the conventional nonlinear operator. First, the delayed Raman response is multiplied by the same signed nonlinear coefficient as the instantaneous Kerr response, although the sign of the Kerr phase shift does not determine the direction of a Stokes transition. Second, the frequency-dependent coefficient $\gamma(\omega)$ multiplies the nonlinear convolution at the generated frequency. In an elementary pump-to-Stokes exchange, the two photon-flux changes therefore contain the factors $\gamma(\omega_\text{p})/\omega_\text{p}$ and $\gamma(\omega_\text{s})/\omega_\text{s}$, respectively. They cancel only for the restricted profile $\gamma(\omega)\propto\omega$. A general nonlinear-dispersion profile breaks this balance~\cite {Linale2020, Vanvincq2011, Zheltikov2018,Linale2020-1}. These are distinct limitations: a signed Raman coefficient can reverse the predicted transfer even when the standard equation conserves photon number. The resulting blueshift and apparent optical energy gain expose the breakdown of this simplified operator. 

Bonetti \textit{et al.} addressed these limitations by deriving a photon-conserving GNLSE (pcGNLSE) in the frequency domain from a quantum description of four-wave mixing and Raman scattering~\cite{Bonetti2020}. As a result, each elementary scattering process 
conserves the total photon number~\cite{Bonetti2019}. Consequently, the pcGNLSE preserves the Stokes-directed Raman frequency transfer for negative Kerr nonlinearity and yields a localized optical-energy decrease whose sign is independent of the Kerr coefficient and the nonlinear-dispersion slope. For positive nonlinearity with $\gamma(\omega)\propto\omega$, the two formulations coincide, confirming its physical consistency in the conventional regime. Note that this equivalence does not result from the linear expansion itself. Rather, it requires the specific relation $\gamma_1=\gamma_0/\omega_0$, corresponding to $\gamma(\omega)\propto\omega$. 
A complementary approach has also studied soliton dynamics with frequency-dependent Kerr nonlinearity in photonic-crystal fibers~\cite{Arteaga2018}. Despite this progress, the frequency-domain formulation is not well suited to direct application of time-domain perturbative and moment methods~\cite{Santhanam2003,Kormokar2024,Liu2025}, which has proven most powerful for studying soliton dynamics analytically. The method of moments requires evolution equations for integral pulse parameters, such as energy, width, and chirp, derived directly in the time domain.

Following up on an earlier investigation by Hernandez {\em et al.} \cite{Hernandez2022}, we derive the time-domain
pcGNLSE through a systematic Taylor expansion of the frequency-dependent photon-conserving nonlinear operator and identify precisely which terms differ from the standard GNLSE. We first apply the method of moments to construct five-parameter evolution systems for bright and dark Raman solitons. We then derive constant-peak-power and constant-blackness attractor conditions for bright and dark solitons, respectively. Finally, we compare representative moment trajectories under the pcGNLSE and the GNLSE. A key finding is that the pcGNLSE produces Raman soliton attractors in both the focusing and defocusing regimes. This
finding opens new possibilities for soliton-based devices
in materials with defocusing nonlinearity.

The paper is organized as follows. Section~\ref{sec:derivation} derives the time-domain pcGNLSE and identifies the additional terms relative to the standard GNLSE. Section~\ref{sec:bright} develops the moment equations and balance condition for bright Raman solitons. Section~\ref{sec:dark} provides the parallel analysis for dark solitons. Section~\ref{sec:simulation} compares representative moment trajectories and reconstructed waveforms from the two models. Section~\ref{sec:conclusions} summarizes the findings.

\section{Time-domain pcGNLSE}
\label{sec:derivation}
\subsection{From the frequency-domain to the time-domain pcGNLSE}
\label{sec:freq}

\noindent In the following, we write $\nu=\omega-\omega_0$ for detuning from the positive carrier angular frequency $\omega_0$. In the notation of Ref.~\cite{Bonetti2020}, the frequency-domain photon-conserving equation is
\begin{equation}
\partial_z\widetilde A(z,\nu)
 =\left[-\frac\alpha2+i\beta(\nu)\right]\widetilde A
 +i\widetilde{\mathcal K}+i\widetilde{\mathcal R}.
\label{eq:freq_pcGNLSE}
\end{equation}
where $\widetilde A (z,\nu)$ is the spectral envelope, $z$ is the propagation distance, $\alpha$ is the linear attenuation coefficient, and $\beta(\nu)=\sum_{n\geq2}\beta_n\nu^n/n!$ is the dispersion relation
Taylor-expanded about the carrier frequency $\omega_0$. With $\mathcal F$ denoting the operator of Fourier transform and $B'=B(T-t')$, nonlinear operators are
\begin{align}
&\widetilde{\mathcal K}=\tfrac12\bar\gamma(\nu)\mathcal F[D^*B^2]
 +\tfrac12\bar\gamma^*(\nu)\mathcal F[B^*D^2],
\label{eq:Kerr_op}\\
&\widetilde{\mathcal R}=f_{\rm R}\bar\gamma^*(\nu)\mathcal F\!\left[
B\!\int_0^\infty h_{\rm R}(t')|B'|^2{\rm d}t'-B|B|^2\right].
\label{eq:Raman_op}
\end{align}
The auxiliary fields are $\widetilde B=k(\nu)\widetilde A$, $\widetilde D=k^*(\nu)\widetilde A$, where $k(\nu)=[\gamma(\omega_0+\nu)/(\omega_0+\nu)]^{1/4}$ and $\bar\gamma(\nu)=[\gamma(\omega_0+\nu)(\omega_0+\nu)^3]^{1/4}$. We consistently choose conjugate roots  across the narrow spectrum around a nonzero $\gamma_0$. The $h_{\rm R}$ in Eq.~\eqref{eq:Raman_op} is the normalized Raman response function, $f_{\rm R}$ is the fractional Raman contribution, and the physical angular frequency is $\omega_0+\nu$. We retain terms through first order in $\nu$. For a real, nonzero coefficient $\gamma_0$, let $\gamma(\omega_0+\nu)=\gamma_0+\gamma_1\nu+O(\nu^2)$ and set $I=|A|^2$. The linearized auxiliary-field factor and outer Kerr factor have the slopes
\begin{equation}
a=\frac14\left(\frac{\gamma_1}{\gamma_0}-\frac1{\omega_0}\right),\qquad
b=\frac14\left(\frac{\gamma_1}{\gamma_0}+\frac3{\omega_0}\right).
\label{eq:nonlinear_slopes}
\end{equation}
Using the detuning operator $\mathcal D\equiv i\partial_t$ with the Fourier convention of Ref.~\cite{Bonetti2020} and $t$ being the delay coordinate, the first-order Kerr operator becomes
\begin{equation}
\mathcal K(\nu)=\gamma_0\bigl[AI+b\mathcal D(AI)\\
 +a\{2I\mathcal DA+A^2(\mathcal DA)^*\}\bigr].
\label{eq:Kerr_expansion}
\end{equation}
Here, higher-order terms $O(\nu^2)$ have been omitted. In particular, $b+3a=\gamma_1/\gamma_0$ and $4\gamma_0a=\gamma_1-\gamma_0/\omega_0$. Hence the Kerr contribution to the time-domain propagation equation is
\begin{equation}
\left.\partial_z A\right|_{\rm K}
 =i\gamma_0AI-\gamma_1\partial_t(AI)
  +\left(\gamma_1-\frac{\gamma_0}{\omega_0}\right)A I_t.
\label{eq:Kerr_time}
\end{equation}
The last term vanishes only for $\gamma_1=\gamma_0/\omega_0$, corresponding locally to $\gamma(\omega)\propto\omega$. The expansion expressed by Eq.~\eqref{eq:Kerr_time} requires $|\nu|\ll\omega_0$ and does not allow for a zero of $\gamma$ within the pulse spectrum.

For a pulse with slow temporal variations compared to the delayed Raman response, we use the first moment $T_{\rm R}$ of that response. Let $R(t)=\int_0^\infty h_{\rm R}(t')I(t-t')\,{\rm d}t'$ and $T_{\rm R}=f_{\rm R}\int_0^\infty t'h_{\rm R}(t')\,{\rm d}t'$. Then
\begin{equation}
f_{\rm R}[R(t)-I(t)]\simeq-T_{\rm R} I_t(t).
\label{eq:Raman_approx}
\end{equation}
The Raman operator must be expanded together with its auxiliary fields. We define $s_\gamma=\gamma_0/|\gamma_0|$ and further $g_+=(s_\gamma\gamma_1+3|\gamma_0|/\omega_0)/4$ and $g_-=(s_\gamma\gamma_1-|\gamma_0|/\omega_0)/4$, which yields the time-domain equation to this order 
\begin{equation}
\begin{split}
A_z={}&-\frac{i\beta_2}{2}A_{tt}+\frac{\beta_3}{6}A_{ttt}
 +i\gamma_0AI\\
&-\gamma_1(AI)_t
 +\left(\gamma_1-\frac{\gamma_0}{\omega_0}\right)AI_t
 -i|\gamma_0|T_{\rm R} AI_t\\
&+T_{\rm R}g_+(AI_t)_t+T_{\rm R}g_-(A_t I_t+IA_{tt}-A^2A_{tt}^*).
\end{split}
\label{eq:pcGNLSE_time}
\end{equation}
where linear loss $\alpha$ is omitted. Equation~\eqref{eq:pcGNLSE_time} contains two independent first-order
expansions. We retain the frequency dependence of the nonlinear operator and the delayed Raman response to first order in the detuning operator $\mathcal D=i\partial_t$ and in the temporal moment $T_{\rm R}$, respectively. 

The derivative terms following the instantaneous Kerr contribution have distinct physical origins. The term $-\gamma_1(AI)_t$ is the usual nonlinear dispersion contribution associated with the slope of $\gamma(\omega)$. The additional term $(\gamma_1-\gamma_0/\omega_0)AI_t$ results from the photon-conserving frequency weighting of the auxiliary fields and vanishes only when $\gamma_1=\gamma_0/\omega_0$. The leading Raman contribution $-i|\gamma_0|T_RAI_t$ is governed by $|\gamma_0|$, and therefore preserves the Stokes direction of the Raman transfer for either Kerr sign. The two terms proportional to $g_+$ and $g_-$ are the corresponding first-order spectral corrections to the Raman operator. Both are required for the photon-conserving expansion and contribute to the energy, momentum, width, and chirp balances derived
below.

\subsection{Normalized form and comparison with the standard GNLSE}
\label{sec:norm}

\noindent Although Eq.~\eqref{eq:pcGNLSE_time} suffices in most cases, we normalize it to facilitate comparison and reduce the number of free
parameters. To this end, we introduce $\xi=z/L_{\rm D}$, $\tau=(T-z/v_{\rm g})/T_0$, $u=A/\sqrt{P_{\rm ref}}$, $L_{\rm D}=T_0^2/|\beta_2|$, $s_{\rm D}=\beta_2/|\beta_2|$, $N^2=|\gamma_0|P_{\rm ref}L_{\rm D}$, $\sigma=\gamma_1P_{\rm ref}L_{\rm D}/T_0$, and $\tau_{\rm R}=T_{\rm R}/T_0$. The normalized field is written as $u=u(\xi,\tau)$. All collective coordinates introduced below are functions of the propagation coordinate $\xi$. Their explicit $\xi$ dependence is omitted for compactness. The reference power $P_{\rm ref}$ is used only in normalizing the field. We define the small carrier-frequency parameter and two Raman coefficients by $\epsilon_\omega=N^2/(\omega_0T_0)$, $\delta=\beta_3/(6|\beta_2|T_0)$, $\kappa_+=(s_\gamma\sigma+3\epsilon_\omega)/4$ and $\kappa_-=(s_\gamma\sigma-\epsilon_\omega)/4$. With $I(\xi,\tau)=|u(\xi,\tau)|^2$, the corrected normalized pcGNLSE reads
\begin{equation}
{\begin{split}
u_\xi={}&-\frac{i s_{\rm D}}2u_{\tau\tau}+\delta u_{\tau\tau\tau}
 +is_\gamma N^2Iu\\
&-\sigma(uI)_\tau+(\sigma-s_\gamma\epsilon_\omega)uI_\tau
 -iN^2\tau_{\rm R}uI_\tau\\
&+\tau_{\rm R}\kappa_+(uI_\tau)_\tau+\tau_{\rm R}\kappa_-(u_\tau I_\tau+Iu_{\tau\tau}
                         -u^2u_{\tau\tau}^*).
\end{split}
\label{eq:normalized_pcGNLSE}}
\end{equation}
The corresponding conventional GNLSE, written with the same $\gamma_0$ and $\gamma_1$, is
\begin{equation}
\begin{split}
u_\xi={}&-\frac{i s_{\rm D}}2u_{\tau\tau}+\delta u_{\tau\tau\tau}
 +is_\gamma N^2Iu-\sigma(uI)_\tau\\
&-is_\gamma N^2\tau_{\rm R}uI_\tau
 +\sigma\tau_{\rm R}(uI_\tau)_\tau.
\end{split}
\label{eq:traditional_GNLSE}
\end{equation}

The comparison includes the carrier-frequency correction to the Kerr term and the full first-order Raman correction, as well as the sign of the leading Raman term. In particular, Eq.~\eqref {eq:normalized_pcGNLSE} allows an arbitrary linear Taylor expansion of $\gamma(\omega)$, which does not require the special relation $\gamma_1=\gamma_0/\omega_0$. For localized bright pulses, the optical-energy balance implied by Eq.~\eqref{eq:normalized_pcGNLSE} is $\partial_\xi\int I\,{\rm d}\tau=-\tau_{\rm R}\epsilon_\omega\int I_\tau^2\,{\rm d}\tau$. The negative sign is independent of $s_\gamma$ and $\sigma$, whereas the corresponding coefficient in the conventional GNLSE is $\sigma$. This difference arises because the photon-conserving Raman correction fixes the energy-transfer term through the positive carrier-frequency scale $\epsilon_\omega$. At the same time, the conventional equation ties the same balance directly to the slope of the nonlinear coefficient. A negative $\sigma$ can therefore produce an optical-energy increase in the conventional model, whereas the pcGNLSE retains a decrease for localized bright pulses. 

\section{Bright Raman soliton attractors}
\label{sec:bright}

\noindent In the regime $s_{\rm D}s_\gamma<0$, i.e., anomalous dispersion with focusing nonlinearity or normal dispersion with defocusing nonlinearity, bright solitons exist. We adopt the method
of moments~\cite{Menyuk1987,Perez-Garcia2007}, defining five integral pulse parameters of energy $E$, time delay $\eta$, frequency shift $\Omega$, pulse width $\widetilde{t}$ and chirp $\widetilde{C}$ as
\begin{align}
&E(\xi)=\int_{-\infty}^{\infty}I(\xi,\tau)\,{\rm d}\tau,
\label{eq:E}\\
&\eta(\xi)=E^{-1}\int_{-\infty}^{\infty}\tau I(\xi,\tau)\,{\rm d}\tau,
\label{eq:eta}\\
&\Omega(\xi)=-E(\xi)^{-1}\int_{-\infty}^{\infty}
                \operatorname{Im}(u^*u_\tau)\,{\rm d}\tau,
\label{eq:Omega}\\
&\widetilde{t}^2(\xi)=E(\xi)^{-1}\int_{-\infty}^{\infty}(\tau-\eta)^2I\,{\rm d}\tau,
\label{eq:t}\\
&\widetilde C(\xi)=-E(\xi)^{-1}\int_{-\infty}^{\infty}
      (\tau-\eta)\operatorname{Im}(u^*u_\tau)\,{\rm d}\tau.
\label{eq:Ctilde}
\end{align}
For the sech trial field, write $x=\tau-\eta$ and
\begin{equation}
u(0,x)=\sqrt{\frac{E}{2\rho}}\,\operatorname{sech}\!\left(\frac{x}{\rho}\right)
 \exp\!\left[i\phi-i\Omega x-\frac{iCx^2}{2\rho^2}\right].
\label{eq:sech}
\end{equation}
\noindent where $\rho$ is the half-width and $C$ is the chirp coefficient. Both are related to the root-mean-square (RMS) quantities by $\widetilde{t}^2=\pi^2\rho^2/12$ and $\widetilde C=\pi^2 C/12$. Substituting Eq.~\eqref{eq:sech} into Eqs.~\eqref{eq:E}--\eqref{eq:Ctilde} and inserting the results into the moment evolution equations yields the following five-coordinate system. Detailed derivations are given in Eqs.~(S29)--(S42) of the Supplemental Material (SM)~\cite{SM}.
\begin{align}
&E'=-\frac{4\tau_{\rm R}\epsilon_\omega E^2}{15\rho^3},
\label{eq:dE_bright}\\
&\eta'=s_{\rm D}\Omega
 +\delta\!\left[3\Omega^2+\frac{1+\pi^2C^2/4}{\rho^2}\right]
 +\frac{E}{3\rho}\left(\frac{\sigma}{2}+s_\gamma\epsilon_\omega\right),
\label{eq:time_delay}\\
&\Omega'=-\frac{4\tau_{\rm R}E}{15\rho^3}
   \left(N^2+s_\gamma\sigma\Omega\right)+\frac{s_\gamma\epsilon_\omega EC}{3\rho^3},
\label{eq:dOmega_bright}\\
&\rho'=\frac{(s_{\rm D}+6\delta\Omega)C}{\rho}
 +\frac{4\tau_{\rm R}\kappa_+ E}{\pi^2\rho^2},
\label{eq:pulse_width}\\
&C'=\frac{s_{\rm D}+6\delta\Omega}{\rho^2}
       \left(C^2+\frac4{\pi^2}\right)
 +\frac{2E(s_\gamma N^2+\sigma\Omega)}{\pi^2\rho}\notag\\
&\quad+\frac{\tau_{\rm R} EC}{\rho^3}
 \left[\frac{3s_\gamma\sigma+7\epsilon_\omega}{\pi^2}
       -\frac{4s_\gamma\sigma}{15}\right].
\label{eq:dC_bright}
\end{align}
\noindent The pulse energy decreases monotonically with propagation, consistent with the physical picture in which Raman scattering transfers photon energy to phonon vibrations. Equation~\eqref{eq:dE_bright} shows that the energy-loss rate is proportional to the carrier-frequency correction $\epsilon_\omega$ and increases strongly as the pulse becomes narrower. In addition, Eq.~\eqref{eq:dE_bright} indicates that energy loss occurs regardless of the sign of the Kerr nonlinear coefficient. The temporal centroid is therefore governed by GVD, TOD, nonlinear dispersion, and the carrier-frequency correction, whose contributions can increase or decrease the temporal shift depending on their signs.

Equation~\eqref{eq:dOmega_bright} contains two distinct contributions. The first term is  Raman-driven, and its leading part, proportional to $N^2$, always drives the frequency toward the Stokes side. The factor $s_\gamma\sigma\Omega$ modifies the Raman shift through nonlinear dispersion. For positive Kerr nonlinearity, a positive $\sigma$ decelerates the soliton self-frequency shift~\cite{Voronin2008}. The second contribution originates from the carrier-frequency Kerr correction and is proportional to $s_\gamma\epsilon_\omega C$. Its sign depends on both the Kerr sign and the chirp. Therefore, the Stokes-directed character of the pcGNLSE is set by the leading Raman contribution rather than by the carrier-frequency correction alone. Comparing Eq.~\eqref{eq:dOmega_bright} with Eq.~\eqref{eq:dE_bright}, one notes that the Raman term governs both energy dissipation and frequency shift, highlighting its intrinsic connection in Raman soliton dynamics. Moreover, Eq.~\eqref{eq:dOmega_bright} indicates that shorter pulses experience a stronger redshift, as all contributions scale as $\rho^{-3}$.

The width evolution contains a dispersive contribution,
$(s_{\rm D}+6\delta\Omega)C/\rho$, and a Raman-derivative contribution
proportional to $\kappa_+=(s_\gamma\sigma+3\epsilon_\omega)/4$.
The latter therefore depends jointly on nonlinear dispersion and the
carrier-frequency correction rather than solely on the Kerr sign. The chirp evolution expressed by Eq.~\eqref{eq:dC_bright} is the most
complex of the five evolution equations because all of the above effects
contribute simultaneously.

Having established the full set of five coupled evolution
equations~\eqref{eq:dE_bright}-\eqref{eq:dC_bright}, we are now in a position to ask which conditions support a self-similar propagation regime, i.e., a soliton attractor. Rather than requiring all five parameters to remain simultaneously fixed, we impose a weaker
constraint, namely, that the peak power $P_\text{pk} \propto E/\rho$ be
preserved. This simplification is justified because the peak power governs the nonlinear phase shift and ultimately determines the soliton character. Since $P_{\rm pk}=E/(2\rho)$ according to Eq.~\eqref{eq:sech}, the resulting criterion for the peak power is
\begin{equation}
\frac{d}{d\xi}\left(\frac E\rho\right)=0.
\label{eq:attractor_def}
\end{equation}
Finally, substituting Eqs.~\eqref{eq:dE_bright} and \eqref{eq:pulse_width} yields
\begin{equation}
\boxed{(s_{\rm D}+6\delta\Omega)C
 +\frac{4\tau_{\rm R} E}{\rho}
 \left(\frac{\epsilon_\omega}{15}
       +\frac{\kappa_+}{\pi^2}\right)=0.}
\label{eq:bright_attractor}
\end{equation}
\noindent Equation~\eqref{eq:bright_attractor} constitutes the central analytical balance condition for bright solitons. The first term describes the combined GVD--TOD contribution, while the second term contains the Raman-induced corrections. An attractor exists whenever these contributions balance. In the absence of the Raman effect ($\tau_{\rm R}$ = 0), the attractor condition reduces to $(s_{\rm D} + 6\delta\Omega)C = 0$, recovering either a chirp-free soliton ($C = 0$) or a zero net-GVD condition $s_{\rm D} + 6\delta\Omega = 0$~\cite{Agrawal2019}.

\section{Dark Raman soliton attractors}
\label{sec:dark}

\subsection{Renormalized moments}
\label{sec:dark_moments}

\noindent When $s_{\rm D}s_\gamma>0$, i.e., for anomalous dispersion with defocusing nonlinearity or normal dispersion with focusing
nonlinearity, dark solitons can form on a finite continuous wave background with nonzero power $P_0$~\cite{Kivshar1998}. The standard
moment definitions diverge because $|u|^2 \to P_0 \neq 0$ as $\tau \to \infty$. We therefore use renormalized moments~\cite{Kivshar1998,Frantzeskakis2010}, replacing $|u|^2$ by the defect density $P_0-|u|^2$ and introducing a regularization factor $\Gamma = 1-P_0/|u|^2$~\cite{Kivshar1994}:
\begin{align}
&E=\int_{-\infty}^{\infty}(P_0-|u|^2)\,{\rm d}\tau,
\label{eq:dark_energy_definition}\\
&\eta=E^{-1}\int_{-\infty}^{\infty}\tau (P_0-|u|^2)\,{\rm d}\tau,
\label{eq:dark_delay_definition}\\
&M=\frac i2\int_{-\infty}^{\infty}
 (uu_\tau^*-u^*u_\tau)\Gamma\,{\rm d}\tau,
\label{eq:dark_frequency_shift_definition}\\
&\widetilde{t}^2=E^{-1}\int_{-\infty}^{\infty}(\tau-\eta)^2(P_0-|u|^2)\,{\rm d}\tau,
\label{eq:dark_width_definition}\\
&\widetilde C=\frac i{2E}\int_{-\infty}^{\infty}
 (\tau-\eta)(u^*u_\tau-uu_\tau^*)\Gamma\,{\rm d}\tau,
\label{eq:dark_chirp_definition}
\end{align}
where $\Gamma=1-P_0/|u|^2$ is a regularization factor that vanishes as $|u|^2-P_0 \to 0$ in the asymptotic limit, ensuring convergence of the momentum integral. The renormalized energy $E$ measures the intensity deficit relative to the continuous-wave background $P_0$. Physically, Eq.~\eqref{eq:dark_energy_definition} quantifies the integrated depth and width of the intensity depletion. The relation $E=2P_0B_{\rm d}^2\rho$ couples the defect energy, blackness, and width. A nearly constant blackness therefore requires the increase of defect energy to be accompanied by a corresponding increase of width. The center-of-mass
coordinate $\eta$ tracks the temporal trajectory of the dark soliton, with $P_0-|u|^2$ serving as the localized weight function. The renormalized momentum $M(\xi)$, weighted by $\Gamma$, is constructed to vanish as $\tau \to \infty$. The evolution of $M$ describes the adiabatic phase dynamics of the soliton and links the Raman perturbation to the spectral shift relative to the background. The renormalized RMS width $\widetilde{t}$ characterizes the temporal extent of the dark soliton, enabling a particle-like description of the soliton. The
chirp $\widetilde{C}$ captures the phase modulation across the soliton
profile and remains well-defined thrthat anhout the einlution as the bkness parameter $B_{\rm d}$ changes.

\subsection{Localized-phase ansatz and evolution equations}
\label{sec:dark_eqs}

\noindent We employ the chirped tanh ansatz~\cite{Kivshar1998}. Retaining the tanh intensity defect, we set $y=(\tau-\eta)/\rho$, $f(y)=\operatorname{sech}^2y$ and use the localized-phase trial field. The blackness parameter $B_{\rm d}$ controls the relative weights of the localized real tanh component and the constant imaginary component while preserving the continuous-wave background intensity $P_0$.
\begin{equation}
\begin{split}
u={}&\sqrt{P_0}\left[B_{\rm d}\tanh y
    +i\sqrt{1-B_{\rm d}^2}\right]\\
&\times\exp\!\left[i\phi_{\rm b}-i\Omega\tau
       +i\Lambda H(y)-iC G(y)\right].
\end{split}
\label{eq:dark_ansatz}
\end{equation}
where $H(y)=-\tanh y\,f(y)/2$ and $G(y)=y\tanh y-\ln\cosh y$, whose derivatives decay at both ends. $B_{\rm d} \in (0, 1]$ is the blackness parameter, with $B_{\rm d} = 1$ corresponding to a black soliton and
$0 < B_{\rm d} < 1$ to a gray soliton. The limit $B_{\rm d}\to0$ removes the localized intensity depression and reduces the ansatz to the uniform continuous-wave background rather than to a dark soliton state. For tanh profiles, the width and chirp relations are $\widetilde{t}^2=\pi^2\rho^2/12$ and $\widetilde{C} = -(\pi^2-6)C/18$. In this section, $\Omega$ denotes the prescribed continuous-wave background frequency and does not drift with $\xi$. The auxiliary coefficient $\Lambda$ represents a localized phase perturbation, which is determined by the momentum $M$ rather than evolved as a sixth independent variable. Substituting Eq.~\eqref{eq:dark_ansatz} into the five definitions in Eqs.~\eqref{eq:dark_energy_definition}--\eqref{eq:dark_chirp_definition} gives,
\begin{align}
&E'=\frac{16\tau_{\rm R}\epsilon_\omega P_0^2B_{\rm d}^4}{15\rho},
\label{eq:dark_dE}\\
&\eta'=s_{\rm D}\eta_1+\delta\eta_2+s_\gamma\epsilon_\omega\eta_3+\sigma\eta_4,
\label{eq:dark_deta}\\
&M'=s_\gamma\epsilon_\omega M_1+\tau_{\rm R}s_\gamma\sigma M_2+\tau_{\rm R}\epsilon_\omega M_3+M_4,
\label{eq:dark_dM}\\
&\rho'=s_{\rm D}\rho_1+\delta\rho_2+\tau_{\rm R}s_\gamma\sigma\rho_3+\tau_{\rm R}\epsilon_\omega\rho_4,
\label{eq:dark_drho}\\
&C'=s_{\rm D}C_1
+\delta C_2
+s_\gamma N^2C_3
+\sigma C_4\\
&\qquad+s_\gamma\epsilon_\omega C_5
+\tau_{\rm R}s_\gamma\sigma C_6
+\tau_{\rm R}\epsilon_\omega C_7, 
\label{eq:dark_dC}
\end{align}
together with $E=2P_0B_{\rm d}^2\rho$. The algebraic relations for $B_{\rm d}$ and $\Lambda$ are given in Eqs.~(S56) and (S57) of the SM. The auxiliary functions $\eta_1$--$\eta_4$, $M_1$--$M_4$, $\rho_1$--$\rho_4$, and $C_1$--$C_7$ are given explicitly in Eqs.~(S64)--(S67), (S74)--(S77), (S82)--(S85), and (S112)--(S118) of the SM, respectively. The blackness parameter $B_{\rm d}$ is not an independent coordinate but is determined algebraically by $E$ and $\rho$. Its evolution follows from the coupled moment equations. Along the balanced pcGNLSE trajectories considered below, $B_{\rm d}$ remains approximately constant. The soliton center trajectory, governed by Eq.~\eqref{eq:dark_deta}, depends on GVD, TOD, the sign of the Kerr nonlinearity, and the SS term. Notably, Eq.~\eqref{eq:dark_dM} shows that neither linear dispersion nor the instantaneous cubic Kerr term contributes directly to $M'$. Instead, the momentum evolution is driven by the carrier-frequency correction and the Raman-related terms. The $B_{\rm d}^4$ dependence in Eq.~\eqref{eq:dark_dE} indicates that the Raman effect is strongest for black solitons and weakens rapidly as the soliton becomes grayer. Equation~\eqref{eq:dark_drho} governs the adiabatic evolution of the pulse width, which is driven by dispersive effects, the Raman effect, and Raman-derivative corrections. Equation~\eqref{eq:dark_dC} reflects the complexity of chirp evolution: dispersion, Kerr nonlinearity, SS, Raman and Raman-SS cross terms contribute simultaneously.

For a dark soliton, the constant-peak-power condition of Sec.~\ref{sec:bright} is replaced by the requirement of constant
blackness, ${\rm d}B_{\rm d}/{\rm d}\xi = 0$, or equivalently, constant peak-power depression. Using $E = 2P_0B_{\rm d}^2\rho$ together with
Eqs.~\eqref{eq:dark_dE} and \eqref{eq:dark_drho}, we obtain the attractor condition for dark solitons:
\begin{equation}
\boxed{B_{\rm d}'=0\quad\Longleftrightarrow\quad
 \rho'=\frac{8\tau_{\rm R}\epsilon_\omega P_0B_{\rm d}^2}{15\rho}.}
\label{eq:dark_attractor_simple}
\end{equation}
Substituting the expanded width expression yields
\begin{equation}
\begin{split}
&-\frac{6s_{\rm D}CH_0}{\pi^2B_{\rm d}^2\rho}
 -\frac{36\delta C}{\pi^2B_{\rm d}^2\rho^2}
 \bigl(\Omega\rho H_0-\Lambda H_1+B_{\rm d}\sqrt{1-B_{\rm d}^2}\,\mathcal J_4\bigr)\\
&=\frac{8\tau_{\rm R}\epsilon_\omega P_0B_{\rm d}^2}{15\rho}-\frac{24\tau_{\rm R}\kappa_+P_0}{\pi^2\rho}
 \left(1-\frac{B_{\rm d}^2}{3}\right).
\end{split}
\label{eq:dark_attractor}
\end{equation}
where $\Lambda$ is given by Eq.~(S57) and $\mathcal J_4$, $H_0$, and $H_1$ are defined in Eqs.~(S60) and (S61) of the SM. Equation~\eqref{eq:dark_attractor} is the dark-soliton counterpart of the bright-soliton attractor condition in Eq.~\eqref{eq:bright_attractor}. The left-hand side collects the dispersive contributions to the width evolution. The right-hand side represents the width change required by constant blackness after the Raman-derivative contribution has been transferred to the same side of the balance. A dark attractor exists when these two sides balance, and this condition can be satisfied for either sign parameter $s_\gamma$ within the corresponding dark-soliton dispersion regime. The dark-soliton balance therefore exhibits the same attractor mechanism as the bright-soliton case, with blackness replacing peak power as the preserved quantity.

\section{Moment-equation results and discussion}
\label{sec:simulation}

\noindent The following results are obtained from the five collective-coordinate equations derived above. The time-domain and frequency-domain waterfalls are reconstructed from the instantaneous variational parameters. The common normalized parameters are $N=1$, $\epsilon_\omega=0.01$, $\tau_{\rm R}=0.05$, and $0\leq\xi\leq10$. The red solid curves denote the pcGNLSE moment system and the blue dashed curves denote the conventional GNLSE moment system.

The normalized parameter sets used below are chosen to expose the qualitative consequences of the photon-conserving corrections rather than to reproduce a specific propagation scenario or material. The carrier-frequency parameter $\epsilon_\omega=N^2/(\omega_0T_0)$ depends on the soliton order, carrier frequency, and pulse duration, while
$\tau_{\rm R}=T_{\rm R}/T_0$ measures the Raman-response time relative to the pulse width. The TOD parameter $\delta=\beta_3/(6|\beta_2|T_0)$ maps directly to the physical dispersion coefficients, whereas $\sigma=\gamma_1P_{\rm ref}L_{\rm D}/T_0$ characterizes the local frequency slope of the nonlinear coefficient. The case-dependent values of $\delta$ and $\sigma$ should therefore be regarded as representative dispersion- and nonlinearity-engineered regimes rather than parameters of a particular material system.

\subsection{Bright Raman soliton trajectories}
\label{sec:bright_sim}

\noindent For all bright cases, $E_0=2$, $\rho_0=1$, $\eta_0=0$, and $\Omega_0=0$. The initial chirp $C_0$ is chosen from Eq.~\eqref{eq:bright_attractor} at $\xi=0$. Table~\ref{tab:bright} lists the case-dependent parameters. The three cases separate a conventional positive-Kerr control, a defocusing case with opposite Raman frequency shifts, and a negative nonlinear-dispersion case with opposite energy trends.

\begin{table}[h]
  \caption{Parameters for the bright-soliton moment calculations.}
  \label{tab:bright}
  \centering
  \begin{tabular}{cccccccc}
    \toprule
    Case\, & $s_{\rm D}$\, & $s_\gamma$\, & $\delta$\, & $\sigma$\,
         & $E_0$\, & $\rho_0$\, & $C_0$ \\
    \midrule
    I   & $-1$ & $+1$ & $-0.0025$ & $+0.01$
        & $2$ & $1$ & $6.72\times10^{-4}$ \\
    II  & $+1$ & $-1$ & $-0.0025$ & $+0.02$
        & $2$ & $1$ & $-3.68\times10^{-4}$ \\
    III & $-1$ & $+1$ & $+0.038$ & $-0.25$
        & $2$ & $1$ & $-1.96\times10^{-3}$ \\
    \bottomrule
  \end{tabular}
\end{table}

\begin{figure}[htbp]
    \centering
    \includegraphics[width=\linewidth]{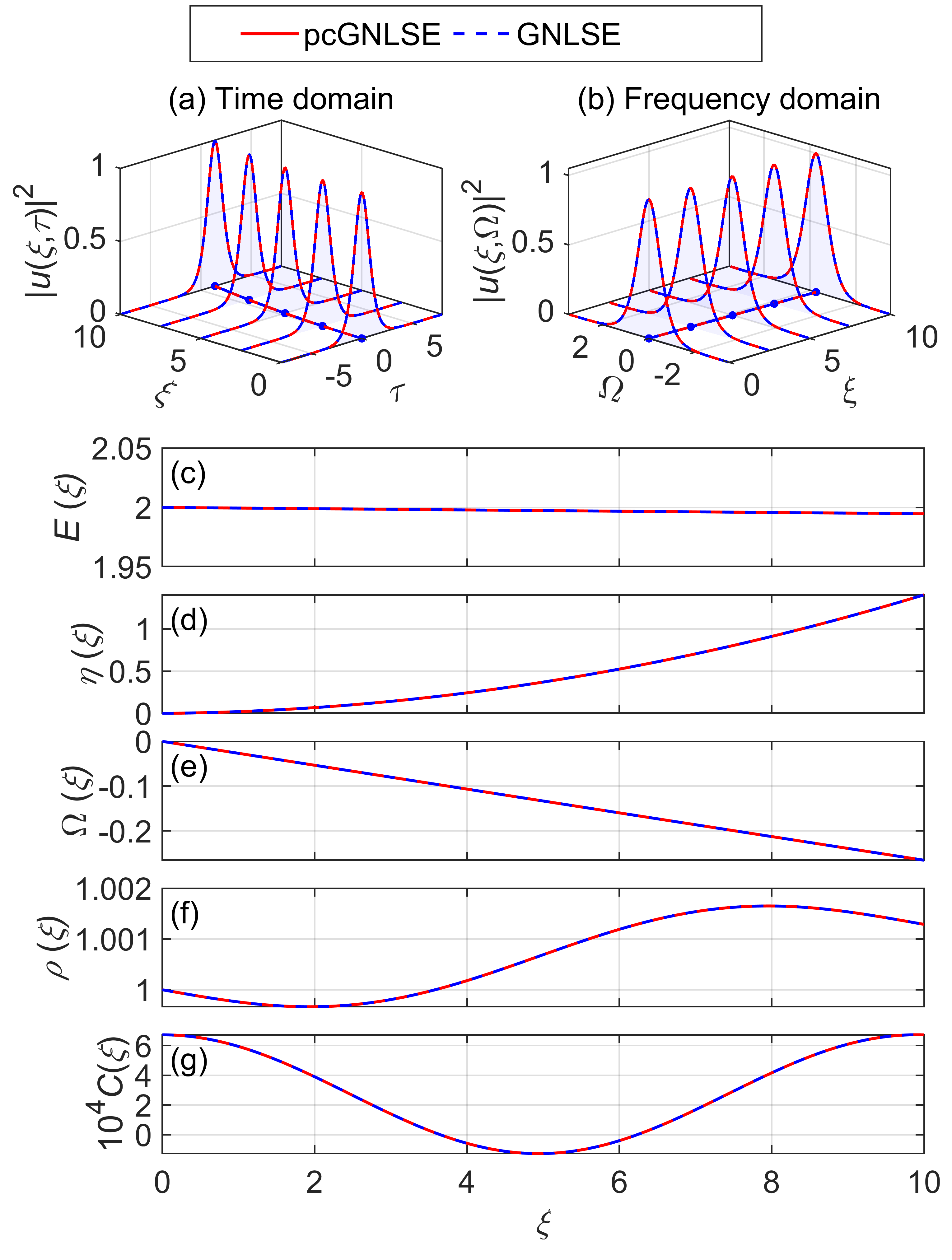}\\[2mm]
   \caption{Bright-soliton moment evolution for Case~I with anomalous dispersion and focusing Kerr nonlinearity. Panels (a) and (b) show the reconstructed temporal intensity and normalized spectrum. Panels (c)--(g) show the pulse energy $E$, temporal centroid $\eta$, mean frequency $\Omega$, pulse width $\rho$, and chirp $C$, respectively. Red solid and blue dashed curves denote the pcGNLSE and the conventional GNLSE, respectively. Colored dots mark the temporal and spectral centers at selected propagation distances. The two models overlap for $\sigma=s_\gamma\epsilon_\omega=0.01$.}
    \label{fig:bright1}
\end{figure}

Case~I satisfies $\sigma=s_\gamma\epsilon_\omega=0.01$, which corresponds
to the local relation $\gamma_1=\gamma_0/\omega_0$. The pcGNLSE and
conventional GNLSE therefore provide an important positive-Kerr control.
Figures~\ref{fig:bright1}(a) and \ref{fig:bright1}(b) show nearly
indistinguishable temporal and spectral evolution. Both trajectories
develop the same temporal displacement and Raman-induced shift toward
lower frequency. Panel~(c) shows a slight decrease in the localized optical
energy for both models. The coincident centroid curves in panel~(d) and
frequency trajectories in panel~(e) confirm that the temporal delay and
Raman redshift are identical in this special limit. The width in panel~(f) remains close to the initial value, consistent with the imposed
constant-peak-power balance. Panel~(g) shows the same small chirp evolution for both models. The agreement across all five collective coordinates provides a direct consistency check for the time-domain pcGNLSE.

\begin{figure}[htbp]
    \centering
    \includegraphics[width=\linewidth]{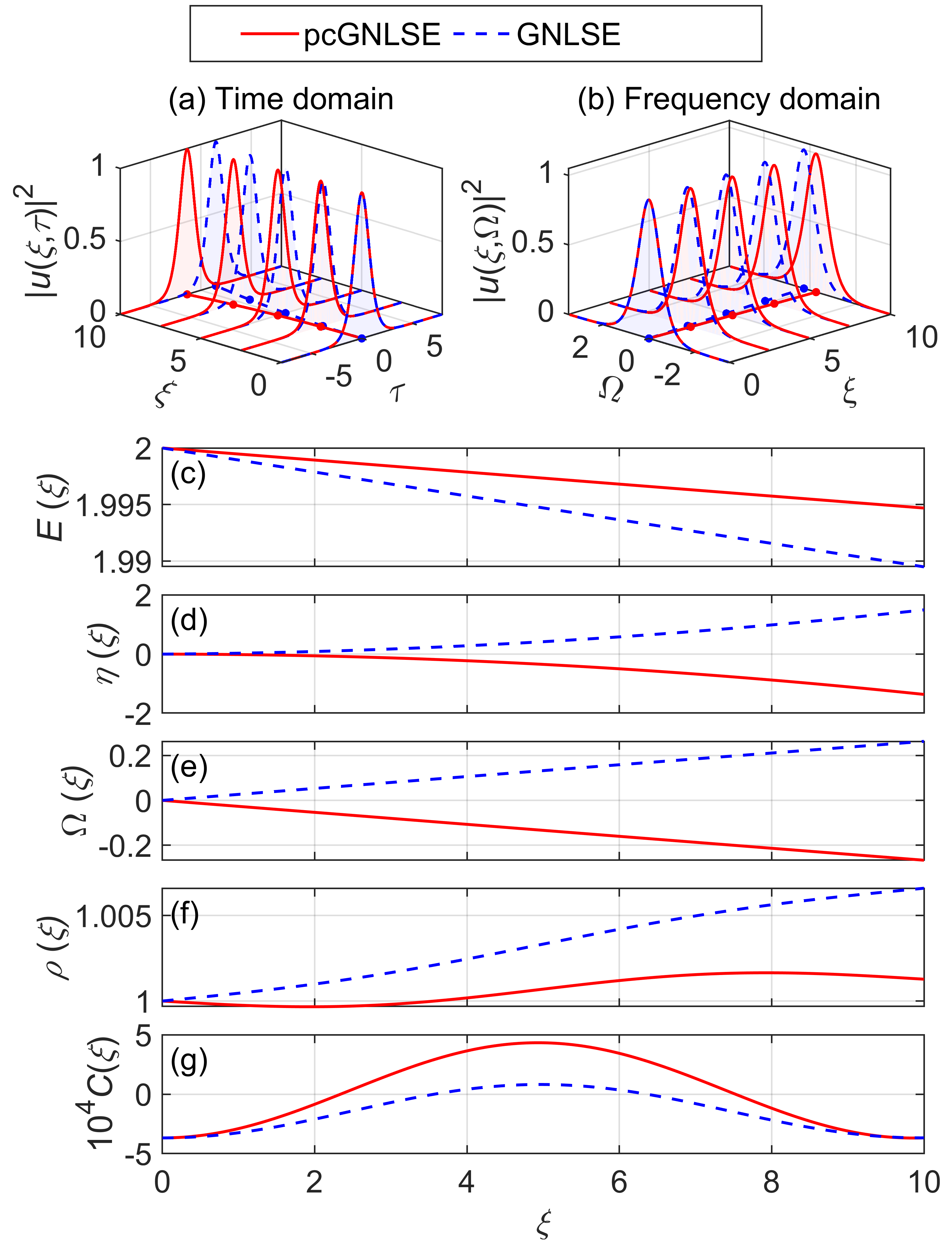}\\[2mm]
   \caption{Bright-soliton moment evolution for Case~II with normal dispersion and defocusing Kerr nonlinearity. Panels and line styles follow Fig.~\ref{fig:bright1}. The pcGNLSE predicts a Raman redshift, whereas the conventional GNLSE predicts a blueshift. The opposite frequency shifts produce opposite temporal drifts and distinct width and chirp evolution.}
    \label{fig:bright2}
\end{figure}

Case~II combines normal dispersion with defocusing Kerr nonlinearity.
The difference between the two models becomes evident in both reconstructed waveforms and collective coordinates. In Fig.~\ref{fig:bright2}(a), the two temporal profiles move in opposite directions as propagation proceeds. The spectral trajectories in panel~(b) separate even more clearly. The pcGNLSE spectrum moves toward lower frequency, whereas the conventional GNLSE spectrum moves toward higher frequency. Panel~(c) shows that the localized optical energy decreases in both models because $\sigma>0$ in the conventional equation. The opposite centroid evolution in panel~(d) follows from the opposite frequency shifts shown in panel~(e). In particular, the pcGNLSE gives $\Omega<0$, while the conventional GNLSE gives $\Omega>0$. Panel~(f) shows that the pcGNLSE width remains close to
the balanced trajectory, whereas the conventional solution undergoes a
larger width change. The chirp trajectories in panel~(g) also separate,
which feeds back into the width evolution through
Eq.~\eqref{eq:pulse_width}. The combined behavior demonstrates that the
sign reversal of the conventional Raman term under negative Kerr
nonlinearity affects not only the spectral shift but the complete
collective-coordinate dynamics.

\begin{figure}[htbp]
    \centering
    \includegraphics[width=\linewidth]{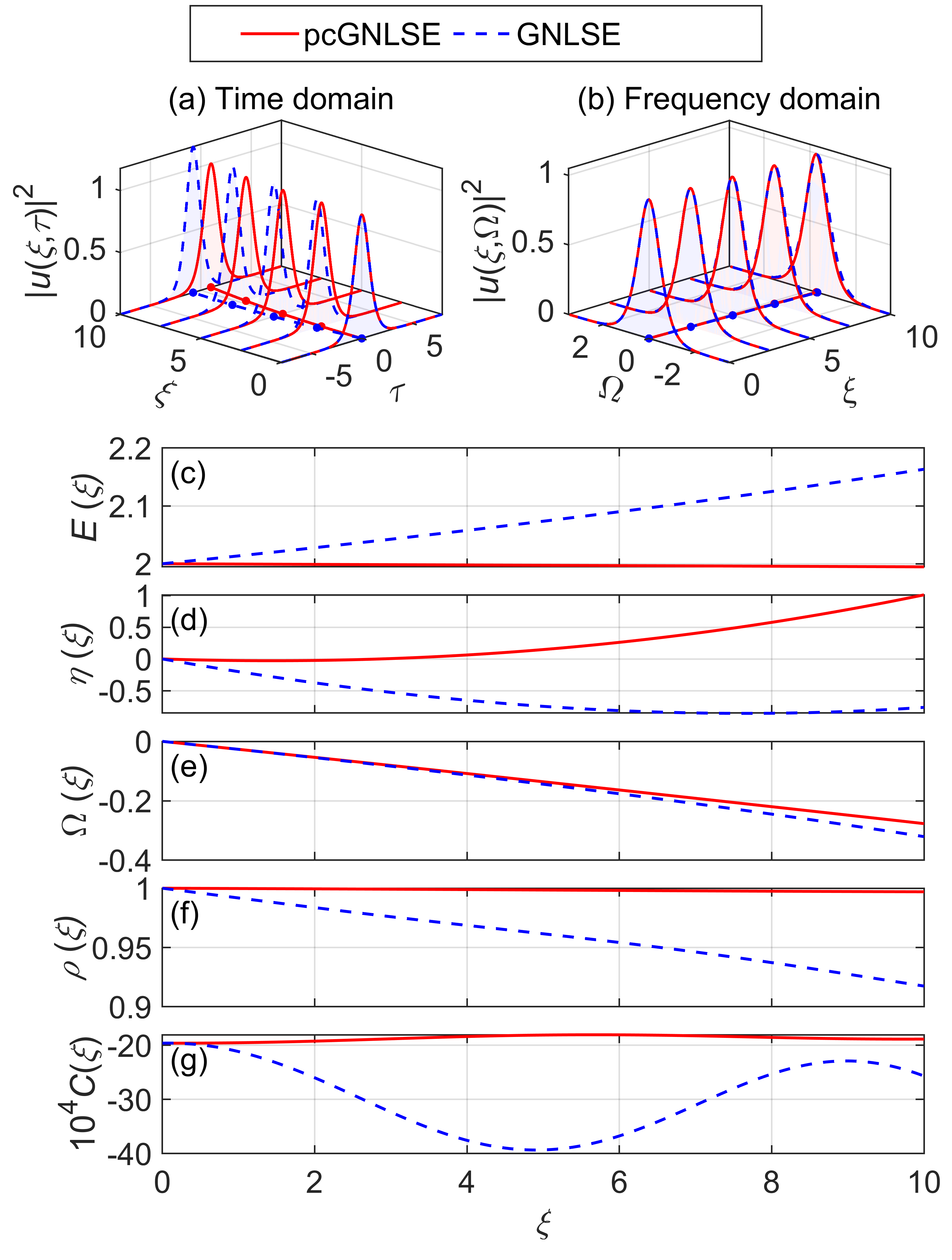}\\[2mm]
    \caption{Bright-soliton moment evolution for Case~III with anomalous dispersion, focusing Kerr nonlinearity, $\delta=0.038$, and $\sigma=-0.25$. Panels and line styles are as in Fig.~\ref{fig:bright1}. Both models shift toward lower frequency, while the energy evolution has opposite signs. The pcGNLSE retains a weak energy decrease and a nearly balanced width, whereas the conventional GNLSE predicts energy growth and pulse narrowing.}
    \label{fig:bright3}
\end{figure}

Case~III retains anomalous dispersion and focusing Kerr nonlinearity while introducing a negative nonlinear-dispersion slope, $\sigma=-0.25$, together with a positive TOD. Figures~\ref{fig:bright3}(a) and
\ref{fig:bright3}(b) show the following separation between the reconstructed
temporal and spectral profiles. In contrast to Case~II, both spectral
trajectories move toward lower frequency. The principal discrepancy
therefore lies elsewhere. Panel~(c) shows opposite energy evolution. The
pcGNLSE predicts a gradual decrease of localized optical energy, whereas
the conventional GNLSE predicts an increase because the conventional
energy-balance coefficient is proportional to $\sigma$. The centroid
trajectories in panel~(d) remain different even though the frequency
shifts in panel~(e) have the same sign. Panel~(f) shows that the pcGNLSE
width stays close to the balanced value, while the conventional trajectory narrows progressively. The corresponding chirp trajectories in panel~(g) also separate and account for part of the different width evolution. Figure~\ref{fig:bright3} therefore isolates the energy-balance failure of the conventional model from the Raman-sign reversal demonstrated in Fig.~\ref{fig:bright2}.

\subsection{Dark Raman soliton trajectories}
\label{sec:dark_sim}

\noindent For all dark-soliton cases, $E_0=1$, $\rho_0=2$, $B_{{\rm d}}=0.99$, $\Omega=0$, and $\Lambda_0=0$. The resulting background power then equals $P_0=E_0/(2B_{{\rm d}}^2\rho_0)\simeq0.255$. The initial chirp $C_0$ is obtained from Eq.~\eqref{eq:dark_attractor} at $\xi=0$. Table~\ref{tab:dark} lists the case-dependent parameters. In this subsection, $E$ denotes the defect energy rather than the total optical energy. A positive derivative in Eq.~\eqref{eq:dark_dE} therefore represents a growth of the intensity void in the continuous-wave background.

\begin{table}[h]
  \caption{Parameters for the dark-soliton moment calculations.}
  \label{tab:dark}
  \centering
  \begin{tabular}{cccccccc}
    \toprule
    Case & $s_{\rm D}$ & $s_\gamma$ & $\delta$ & $\sigma$
         & $\rho_0$ & $B_{{\rm d},0}$ & $C_0$ \\
    \midrule
    I   & $+1$ & $+1$ & $-0.0025$ & $+0.01$
        & $2$ & $0.99$ & $1.87\times10^{-4}$ \\
    II  & $-1$ & $-1$ & $+0.035$ & $+0.20$
        & $2$ & $0.99$ & $1.26\times10^{-3}$ \\
    III & $-1$ & $-1$ & $-0.020$ & $-0.20$
        & $2$ & $0.99$ & $-1.49\times10^{-3}$ \\
    \bottomrule
  \end{tabular}
\end{table}

\begin{figure}[htbp]
    \centering
    \includegraphics[width=\linewidth]{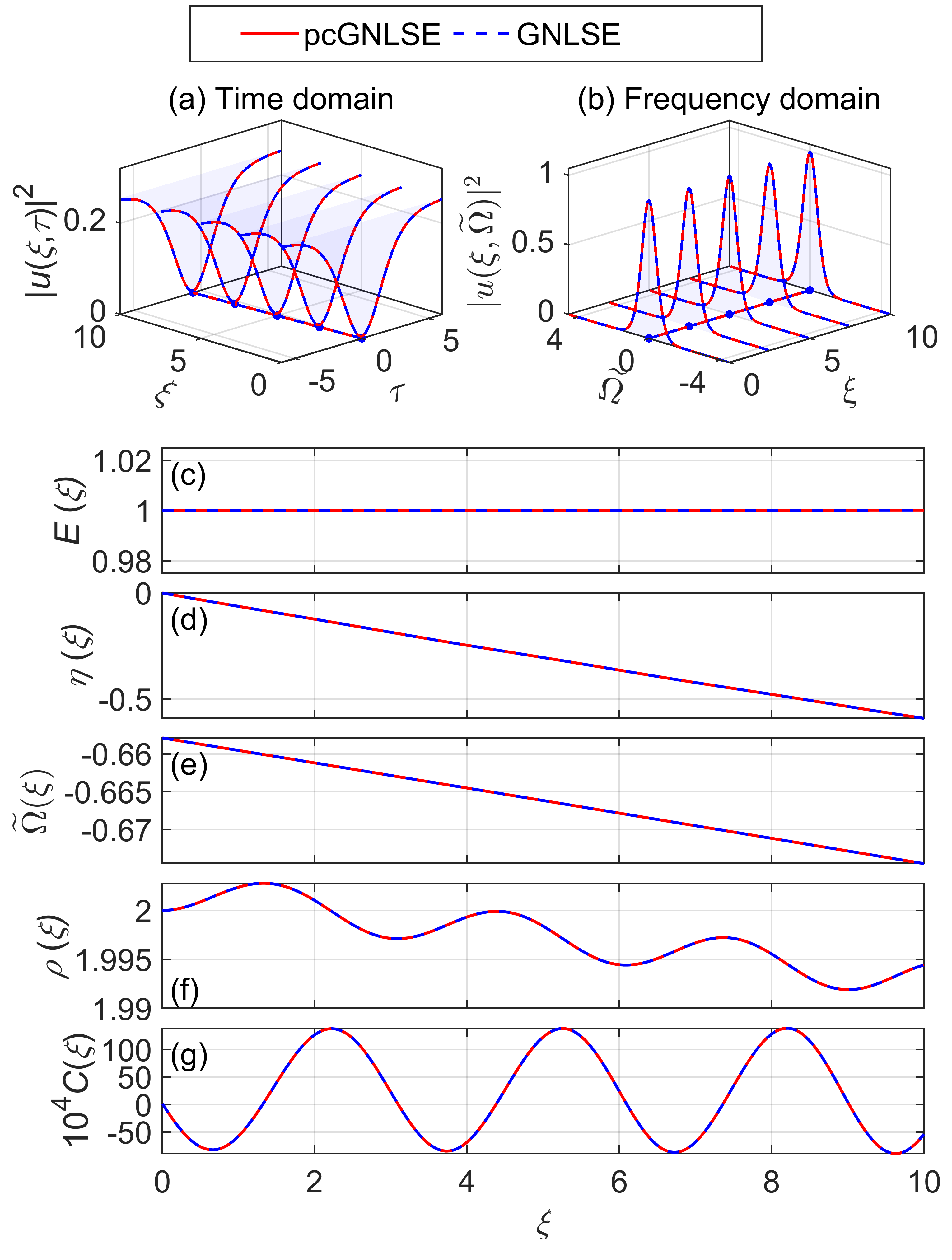}\\[2mm]
   \caption{Dark-soliton moment evolution for Case~I with normal dispersion and focusing Kerr nonlinearity. Panel (a) shows the reconstructed dark-notch intensity, with shading restricted to the localized intensity deficit. Panel (b) shows the localized spectral waterfall. Panels (c)--(g) show the defect energy $E$, temporal centroid $\eta$, renormalized frequency $\widetilde{\Omega}=-M/E$, pulse width $\rho$, and chirp $C$, respectively. Red solid and blue dashed curves denote the pcGNLSE and the conventional GNLSE, respectively. Colored dots mark the temporal and spectral centers at selected propagation distances. The two models overlap for $\sigma=s_\gamma\epsilon_\omega=0.01$.}
    \label{fig:dark1}
\end{figure}

Case~I provides the dark-soliton counterpart of the positive-Kerr control. The condition $\sigma=s_\gamma\epsilon_\omega=0.01$ again makes the two reduced models coincide. Figure~\ref{fig:dark1}(a) shows the reconstructed dark notch on the finite continuous-wave background. The shaded region represents only the localized intensity deficit. Panel~(b) shows identical localized spectral evolution in both models. The defect energy in panel~(c) increases weakly as it propagates. This positive trend does not represent an increase in total optical energy, since $E$ measures the area of the intensity void relative to the background. The overlapping centroid curves in panel~(d) show the same temporal trajectory. Panel~(e) shows the common evolution of the renormalized frequency $\widetilde\Omega=-M/E$. The width in panel~(f) increases slightly together with the defect energy, while the nearly unchanged blackness follows from the balance between $E$ and $\rho$. The chirp trajectories in panel~(g) remain identical. Agreement across all seven panels confirms the expected equivalence of the two models in this positive-Kerr special case.

\begin{figure}[htbp]
    \centering
    \includegraphics[width=\linewidth]{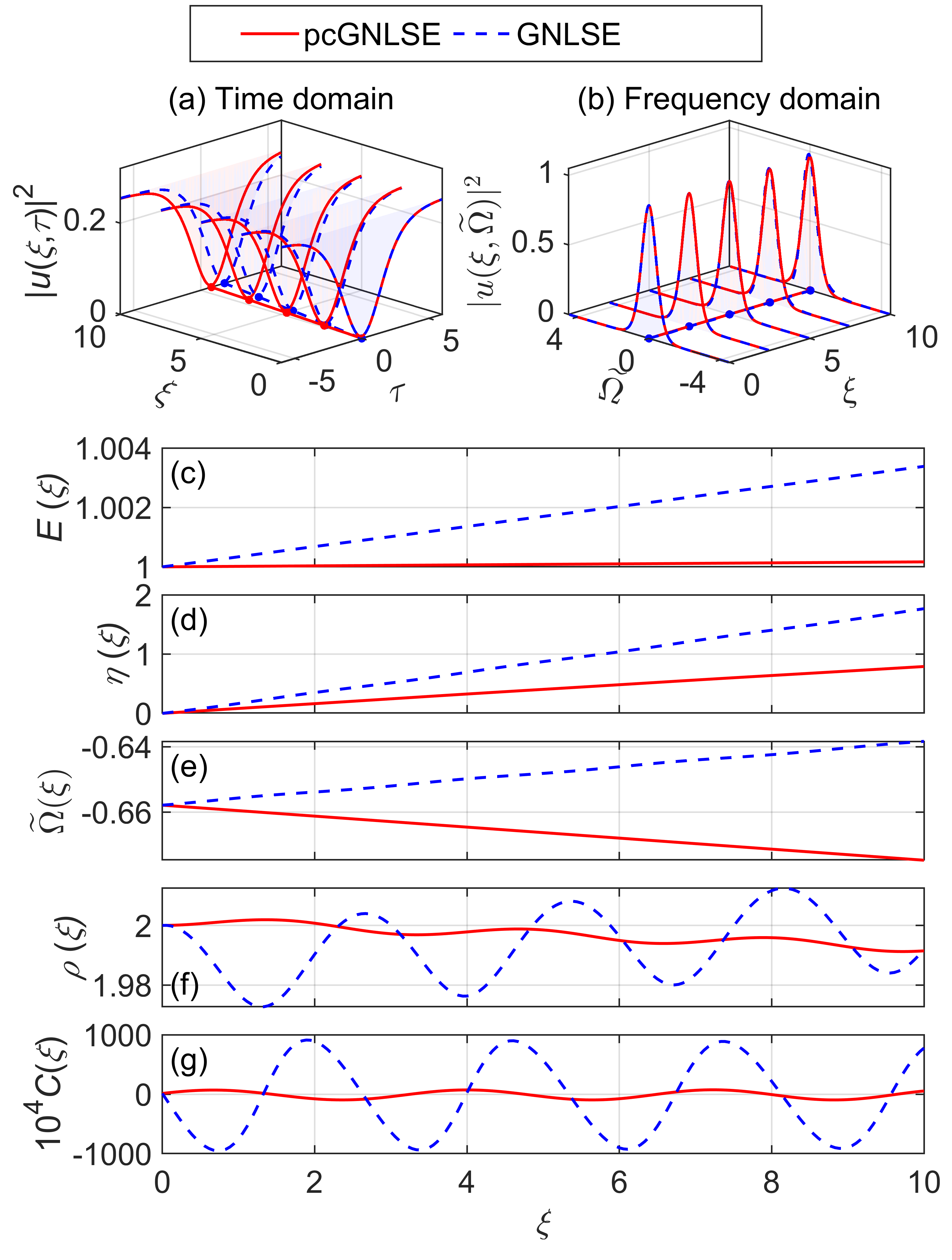}\\[2mm]
    \caption{Dark-soliton moment evolution for Case~II with anomalous dispersion, defocusing Kerr nonlinearity, $\delta=0.035$, and $\sigma=0.20$. Panels and line styles are as in Fig.~\ref{fig:dark1}. The two models develop distinct temporal and spectral trajectories, together with pronounced differences in defect energy, width, and chirp. The pcGNLSE remains close to the constant-blackness trajectory.}
    \label{fig:dark2}
\end{figure}

Case~II changes both the dispersion and Kerr signs and introduces a
positive nonlinear-dispersion slope, $\sigma=0.20$. The two models now
produce clearly different dark-soliton dynamics. In
Fig.~\ref{fig:dark2}(a), the reconstructed notches develop different
temporal trajectories while the localized intensity depression remains
nearly preserved for the pcGNLSE. Panel~(b) shows a corresponding
separation of the localized spectra. The defect energies in panel~(c)
increase in both models, although the conventional GNLSE produces a much
greater change. The centroid trajectories in panel~(d) separate
progressively, consistent with the different renormalized-frequency
evolution in panel~(e). The pcGNLSE frequency coordinate changes smoothly,
whereas the conventional trajectory develops a substantially different
shift. Panel~(f) reveals an even stronger contrast in the width dynamics.
The pcGNLSE remains close to the constant-blackness trajectory, while the
conventional width exhibits much larger excursions. The chirp in panel~(g)
shows the same qualitative distinction and develops stronger variation in
the conventional model. The correlated changes of $\rho$ and $C$ explain
why the waveform difference in panel~(a) becomes visible even though the
pcGNLSE blackness remains nearly constant.

\begin{figure}[htbp]
    \centering
    \includegraphics[width=\linewidth]{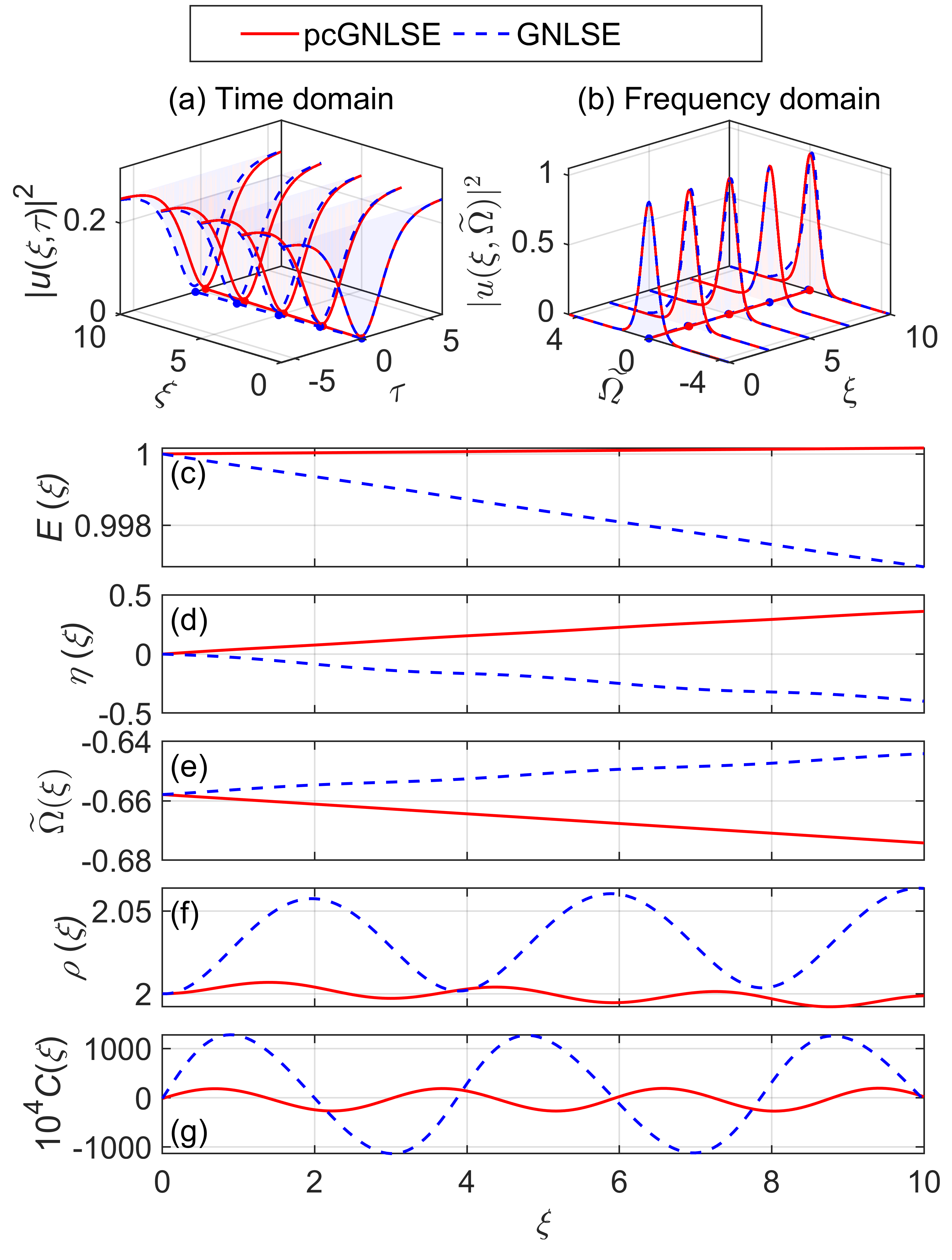}\\[2mm]
   \caption{Dark-soliton moment evolution for Case~III with anomalous dispersion, defocusing Kerr nonlinearity, $\delta=-0.020$, and $\sigma=-0.20$. Panels and line styles are as in Fig.~\ref{fig:dark1}. The pcGNLSE predicts a weak increase of defect energy, whereas the conventional GNLSE predicts a decrease. Opposite centroid and renormalized-frequency drifts are accompanied by distinct width and chirp evolution.}
    \label{fig:dark3}
\end{figure}

Case~III retains anomalous dispersion and defocusing Kerr nonlinearity but
reverses the nonlinear-dispersion slope to $\sigma=-0.20$. The TOD
coefficient is also changed to $\delta=-0.020$. The reconstructed
time-domain profiles in Fig.~\ref{fig:dark3}(a) separate increasingly with
propagation and move along different temporal trajectories. Panel~(b)
shows a simultaneous separation of the localized spectral distributions.
A particularly clear distinction appears in panel~(c). The pcGNLSE defect
energy increases weakly, in agreement with Eq.~\eqref{eq:dark_dE}, whereas
the conventional GNLSE predicts a decrease. The centroid trajectories in
panel~(d) develop opposite signs. The renormalized-frequency trajectories
in panel~(e) also move in opposite directions. Panel~(f) shows that the
two models predict markedly different width evolution. The difference is
accompanied by strongly separated chirp trajectories in panel~(g). The
combined behavior shows that reversing the nonlinear-dispersion slope
changes the conventional energy trend and reorganizes the full dark-soliton
dynamics, while the sign of the pcGNLSE defect-energy balance remains
fixed by $\epsilon_\omega>0$.   

\section{Conclusions}
\label{sec:conclusions}

\noindent
We have derived a first-order time-domain photon-conserving GNLSE from the established frequency-domain formulation and used it to construct a moment description of Raman soliton dynamics. The photon-conserving expansion introduces carrier-frequency corrections into both the instantaneous Kerr and Raman operators. For localized bright pulses, the pcGNLSE predicts an optical-energy decrease governed by the positive parameter $\epsilon_\omega$, independently of the signs of the Kerr coefficient and the nonlinear dispersion slope. For dark solitons, the corresponding defect-energy balance is positive because the background-subtracted quantity measures the intensity void due to the dark soliton. The leading Raman contribution also retains the Stokes-directed photon transfer when the Kerr nonlinearity is defocusing, thereby avoiding the sign reversal predicted by the conventional GNLSE.

Using the method of moments, we obtained closed evolution equations for five collective coordinates describing both bright and dark solitons. The resulting constant-peak-power and constant-blackness conditions identify Raman-soliton attractors independent of the nonlinearity sign. For bright solitons, the balance condition yields an explicit relation among chirp, frequency shift, dispersion, and Raman-induced corrections. For dark solitons, the corresponding condition links chirp, width, blackness, background power, and localized phase contribution. These relations show that the photon-conserving Raman corrections do more than modify individual trajectories. They determine whether a self-similar balance can be maintained when the Kerr sign or nonlinear-dispersion slope is reversed. The analytical conditions can therefore be used in reverse to identify pulse and material parameters that support a desired balanced trajectory.

Integrated photonic platforms increasingly employ engineered or intrinsically negative-$\chi^{(3)}$ nonlinearities, including two-dimensional materials~\cite{Vermeulen2016,Dremetsika2016}, metamaterials~\cite{Zhang2017}, and semiconductor waveguides~\cite{A-P2025}. In such systems, the conventional GNLSE can give qualitatively incorrect Raman frequency or energy evolution outside the usual positive-Kerr regime. The pcGNLSE framework developed here provides a physically consistent analytical basis for designing Raman-soliton dynamics in these less conventional nonlinear media. Extension of the present attractor picture to multimode systems, dispersion-engineered waveguides, and microresonators may provide further routes toward deterministic control of Raman-shifted soliton states.

\begin{acknowledgments}
    \noindent This work was partially supported by the National Natural Science Foundation of China (62275015 and 62205015) and the National Key R$\&$D Program of China (2024YFB3613503).
\end{acknowledgments}

\makeatletter\immediate\write\@auxout{\string\citation{pcGNLSEControl}}\makeatother
\bibliographystyle{apsrev4-2}
\bibliography{Reference}

\end{document}